\documentstyle[twocolumn,aps,prl,psfig]{revtex}
\tighten
\draft
\begin{document}

\draft
\twocolumn[\hsize\textwidth\columnwidth\hsize\csname@twocolumnfalse\endcsname

\preprint{\today}
\title{How does the relaxation of a supercooled liquid depend on its
microscopic dynamics?}
\author{Tobias Gleim, Walter Kob, and Kurt Binder}
\address{Institut f\"ur Physik, Johannes Gutenberg-Universit\"at,
Staudinger Weg 7, D-55099 Mainz, Germany}
\maketitle

\begin{abstract}
Using molecular dynamics computer simulations we investigate how
the relaxation dynamics of a simple supercooled liquid with Newtonian
dynamics differs from the one with a stochastic dynamics. We find
that, apart from the early $\beta$-relaxation regime, the two dynamics
give rise to the same relaxation behavior. The increase of the
relaxation times of the system upon cooling, the details of the
$\alpha$-relaxation, as well as the wave vector dependence of the
Edwards-Anderson-parameters are independent of the microscopic
dynamics.

\begin{center} \today \end{center}
\end{abstract}

%\twocolumn

\vskip2pc]

\narrowtext
\noindent
If the logarithm of a transport quantity, such as the viscosity $\eta$,
of a good glass former is plotted versus $T_g/T$, where $T$ is
temperature and $T_g$ is the glass transition temperature, it becomes
obvious that this temperature dependence is not universal since some
materials show essentially an Arrhenius behavior whereas others show
a strong non-Arrhenius behavior~\cite{angell85}. Also more microscopic
dynamical properties, such as the Raman spectrum, depend strongly on
the material, in that, i.e., the so-called boson peak is much more
pronounced in strong glass formers than in fragile glass
formers~\cite{rossler94}. Thus we can say that it is well established
that the macroscopic as well as the microscopic dynamics of supercooled
liquids is not universal at all and must be considered as a material
specific property. This insight is of course not surprising, since
the materials differ in their structure, the masses of the individual
atoms etc. and thus it can be expected that these microscopic
quantities will give rise to a different relaxational and vibrational
dynamics. 

What is much less obvious, however, is how the {\it microscopic
dynamics} affects the vibrational and relaxational dynamics of the
system, i.e. whether the relaxational dynamics is different if the
microscopic dynamics is, e.g., a Newtonian one or a Brownian one. The
answer to this question is most important since it will allow to gain
insight to understand which aspects of the relaxation behavior are, for
a given system, universal and which ones are not. This information is in
turn relevant for testing the applicability of theories that attempt
to describe the slowing down of the system upon cooling, i.e. the
mechanism for the glass transition.

In real experiments it is of course difficult to investigate how the
microscopic dynamics affects the dynamics of the system at long times,
since usually it is not possible to change the former without also
influencing other microscopic quantities like, e.g., the masses of the
particles or the interaction between the atoms. Experimentally a
Brownian type dynamics can be realized, e.g., by colloidal
fluids~\cite{colloids}, while atomic liquids have a Newtonian
dynamics. However, the structure and interparticle forces in these two
types of systems are quite distinct from each other and thus it is not
surprising that the two corresponding dynamics are different.  For
computer simulations it is, however, most simple to change, for a given
system, the dynamics and they are therefore ideally suited to
investigate such questions.  The only investigation in that direction
we know of is a pioneering study by L\"owen {\it et
al.}~\cite{lowen91}. In that work the authors compared the relaxation
dynamics of a polydisperse system of charged particles which move
according to a Newtonian dynamics to the one moving according to a
Brownian dynamics. The outcome of that study was that the relaxation
dynamics depends, on the time scale of the
$\beta$-relaxation, on the microscopic dynamics.
However, due to the limited length of the runs and the lack of
statistics no stringent test could be made whether or not the
$\alpha$-relaxation depends on the microscopic
dynamics. If such a dependence would also exist in the
$\alpha$-relaxation regime it would be in contrast to the prediction of
the so-called mode-coupling theory (MCT)~\cite{mct1,mct2}, according to
which the relaxation dynamics at long times should be independent of
the microscopic dynamics. Since there is very good evidence that this
theory gives a reliable description of the dynamics of supercooled
simple liquids and even network forming liquids,
see~\cite{barrat91,kob,sciortino} and references therein, such a
disagreement between theory and computer simulation would be quite
disturbing since it shows that certain {\it fundamental} aspects of the
theory are not correct.  Due to the availability of better computers
and algorithms it is today possible to do simulations which are more
than hundred times longer than the ones done by L\"owen {\it et al.}
Therefore it is possible to investigate the question how the long time
dynamics of a system depends on the microscopic dynamics on a
qualitatively completely different level and in the present paper we
report the results of such an investigation.

The system considered is a 80:20 mixture of 1000 Lennard-Jones
particles consisting of two species of particles, $A$ and $B$.  All the
particles have the same mass $m$ and the interaction between two
particles of type $\alpha, \beta \in \{A,B\}$ is given by
$V_{\alpha\beta}(r)=4\epsilon_{\alpha\beta}
[(\sigma_{\alpha\beta}/r)^{12} -(\sigma_{\alpha\beta}/r)^6]$ with
$\epsilon_{AA}=1.0$, $\sigma_{AA}=1.0$, $\epsilon_{AB}=1.5$,
$\sigma_{AB}=0.8$, $\epsilon_{BB}=0.5$, and $\sigma_{BB}=0.88$, with a
cut-off radius of $2.5\sigma_{\alpha\beta}$. In the following we will
always use reduced time units with $\sigma_{AA}$ and $\epsilon_{AA}$
the unit of length and energy, respectively (setting the Boltzmann
constant $k_B$ equal to 1.0). Time is measured in units of
$\sqrt{\sigma_{AA}^2m/48\epsilon_{AA}}$. The volume of the box is kept
constant with a box length of 9.4. Two types of microscopic dynamics
are investigated: A Newtonian one and a stochastic one (described
below). In the Newtonian dynamics (ND) we integrate Newton's equation of
motion with the velocity form of the Verlet algorithm with a time step
of 0.02. After equilibrating the system in the canonical ensemble
we turn off the heat bath and start the production run in the
microcanonical ensemble. For this kind of dynamics the
temperature dependence of the relaxation behavior has already been
studied in great detail in the temperature range $5.0 \geq T \geq
0.466$~\cite{kob}.  Due to an improvement of the
hardware and computer codes it is, however, possible today to perform
simulations which extend over a time range that exceed the ones of the
previous investigations by a factor of eight, giving now a total
of $4\times 10^7$ time steps, and thus allowing us to equilibrate the
system at lower temperatures than has been possible before and hence to
perform more accurate tests. Therefore we have determined the dynamics
of the system also at $T=0.452$ and 0.446.

The stochastic dynamics (SD) we considered is defined as follows:
Apart from the deterministic forces that originate from the interaction
potential given above, each particle $j$ is also subject to a gaussian
distributed white noise force $\vec{\eta}_j(t)$ with zero mean, i.e.,
$\langle \vec{\eta}_j(t)\rangle =0$, and a damping force which is
proportional to the velocity of the particle.  Thus the equation of
motion for particle $j$ reads:
\begin{equation}
m\ddot{\vec{r}}_j+\nabla_j \sum_l V_{\alpha_j\beta_l}(|\vec{r}_l-\vec{r}_j|)
=-\zeta \dot{\vec{r}}_j + \vec{\eta}_j(t)\quad,
\label{eq1}
\end{equation}
where the damping constant $\zeta$ is given from the fluctuation
dissipation theorem, i.e. $\langle \vec{\eta}_j(t)\cdot
\vec{\eta}_l(t') \rangle = 6k_B T \zeta \delta(t-t')$. The value of
$\zeta$ was set to 10, which is sufficiently large that the presented
results for the SD do not depend on $\zeta$ anymore (apart from a
trivial change of the time scale). Equations~(\ref{eq1}) where solved
with a Heun algorithm~\cite{paul95}. The time step used was 0.008,
which is small enough to ensure that the {\it equilibrium} properties
of the system are the same for the ND and SD (this was checked at all
temperatures). Also for this dynamics the runs at the lowest
temperature extended over $4\times 10^7$ time steps. At the lowest
temperatures the SD is so slow that, within the time span of the runs,
the system does not equilibrate. Therefore we used at each temperature
the ND to equilibrate the system and used the SD only for the
(equilibrium) production runs. In order to improve the statistics of
the results we averaged at each temperature over eight independent runs
(ND as well as SD). The temperatures investigated were 5.0, 4.0, 3.0,
2.0, 1.0, 0.8, 0.6, 0.55, 0.500, 0.475, 0.466, 0.452, and 0.446.

\begin{figure}[f]
\psfig{file=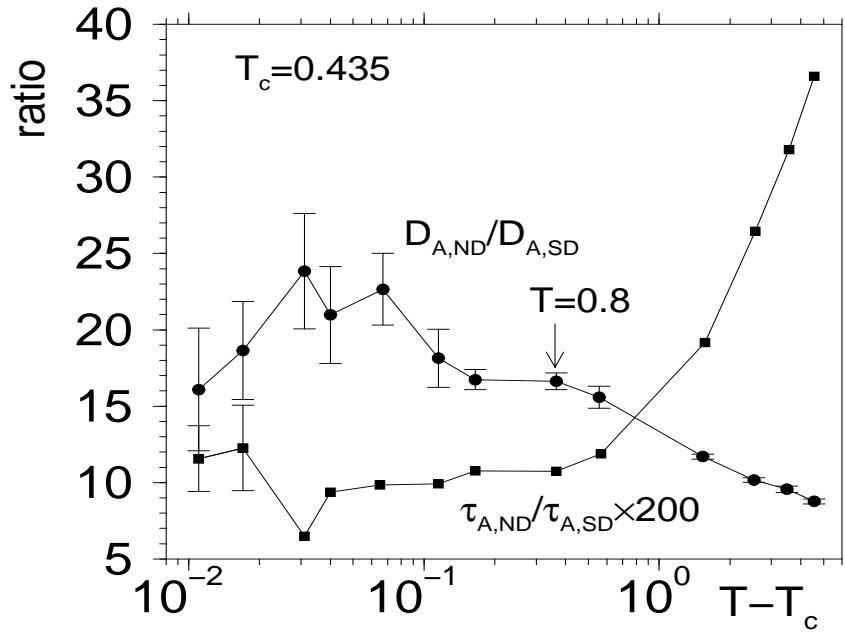,width=13cm,height=9.5cm}
\caption{Temperature dependence of the ratio of the diffusion constant 
for the $A$ particles for the ND and SD (circles).
Squares: Ratio for the $\alpha$-relaxation time of the incoherent 
intermediate scattering function.}
\label{fig1}
\end{figure}

One of the simplest quantities to characterize the dynamics of the
system is the diffusion constant $D_{\alpha}$ of the particles, which
we calculated from the mean squared displacement of a tagged particle.
For the ND the temperature dependence of $D_{\alpha}$ has been
determined before~\cite{kob} and it was shown that, at low
temperatures, it is given by a power-law, i.e. $D_{\alpha} \propto
(T-T_c)^{\gamma}$, a functional form predicted by MCT. (Here $T_c>0$ is
the so-called critical temperature of MCT.) It is, however, desirable
to compare the temperature dependence of $D_{\alpha}$ for the ND with
the one for the SD without making reference to any theory and thus we
calculated the ratio $D_{A,ND}/D_{A,SD}$ and show its temperature
dependence in Fig.~\ref{fig1}. Note that we plot this ratio versus
$T-T_c$, were $T_c=0.435$ is the critical temperature from the MCT
analysis~\cite{kob}, instead of $T$, but we emphasize that this
representation of the data has nothing to do with MCT but is only a
conveniant way to expand the abscissa at low temperatures. From this
figure we see that this ratio shows a noticable temperature dependence
at high temperatures but becomes essentially constant (within the error
bars) for temperatures $T \leq 0.8$. We note that in the temperature
interval $0.446 \leq T \leq 0.8$ the diffusion constants change by
almost three decades~\cite{gleim98}. Since the ratio
$D_{A,ND}/D_{A,SD}$ stays constant to within about 30\% we conclude
that the temperature dependence of the diffusion constants is
independent of the microscopic dynamics to within a few parts in
$10^4$. A similar result is found for the $B$ particles.

\begin{figure}[h]
\psfig{file=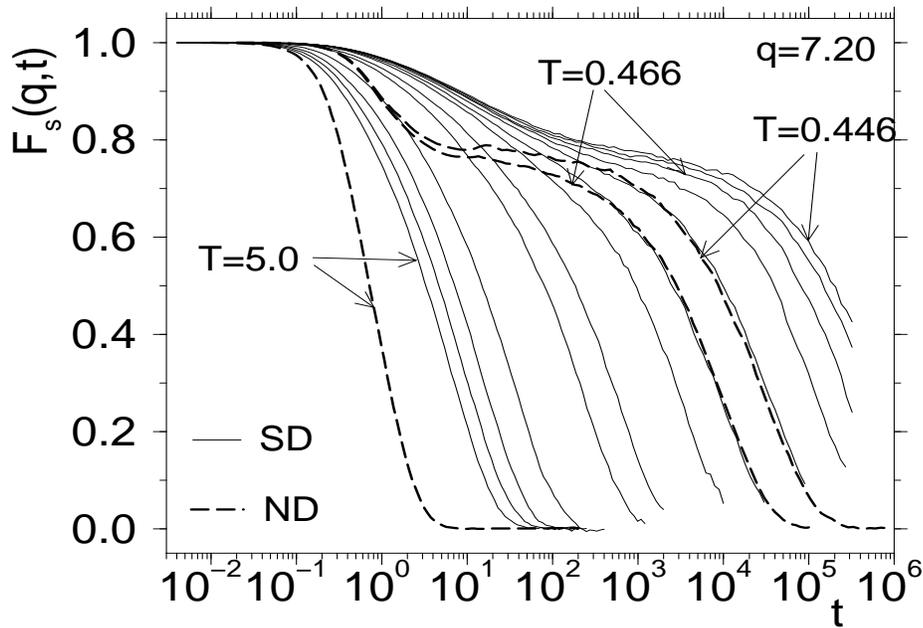,width=13cm,height=9.5cm}
\caption{Time dependence of the incoherent intermediate scattering function
for the SD for all temperatures investigated (solid lines) and the ND for
selected temperatures (dashed lines).}
\label{fig2}
\end{figure}

A further quantity that is very useful to characterize the dynamics of
the system is the (incoherent) intermediate scattering function
$F_s(q,t)$ for wave vector $q$~\cite{hansen_mcdonald86}. For the SD we
show the time dependence of this quantity in Fig.~\ref{fig2} for all
temperatures investigated (solid lines). The value of $q$ is 7.20, the
maximum of the static structure factor for the $A-A$
correlation~\cite{kob}.  It can be seen that on lowering the
temperature the relaxation behavior changes qualitatively in that at
high temperatures the decay of $F_s(q,t)$ is essentially an exponential
(apart from the time dependence at very short times) whereas it shows a
plateau at low temperatures. The time for which the correlation
function decays to zero increases quickly with decreasing temperature
(note the logarithmic time axis) indicating the dramatic slowing down
of the relaxation dynamics of the system upon cooling. This strong
dependence of the dynamics on temperature has also been seen in the
case of the ND~\cite{kob}. We emphasize that these curves are all {\it
equilibrium} curves.

Also shown in the figure is $F_s(q,t)$ for the ND at three different
temperatures (dashed curves). We first compare the SD and ND in the
$\alpha$-relaxation regime, i.e. the time regime where the correlation
function decays, at low temperatures, from the mentioned plateau. From
the figure we see that at the highest temperature ($T=5.0$) the ND
gives rise to a relaxation that is about a factor of seven faster than
the one for the SD. This factor increases upon cooling the system and
reaches about 20-30 at low temperatures. Thus from the point of view of
the {\it absolute values} of the $\alpha$-relaxation times $\tau(T)$,
the two types of dynamics are very different. ($\tau(T)$ can be
defined, e.g., as the time the correlator takes to decay to $e^{-1}$ of
its initial value.) However, if we look at the {\it temperature
dependence} of $\tau(T)$, we come to a different conclusion. In
Fig.~\ref{fig1} we also show the ratio $\tau_{A,ND}/\tau_{A,SD}$ and find
that this ratio becomes independent of $T$ for temperatures smaller
than 1.0. Thus the ratio shows a similar dependence on temperature as
the one for the diffusion constants. 

From Fig.~\ref{fig2} we also see that the shape of the correlation
functions {\it in the $\alpha$-relaxation regime} is independent of the
microscopic dynamics. This can be recognized by the fact that the curve
for at $T=0.466$ for the ND lies almost on top of the curve for the SD
for $T=0.55$ {\it and} the fact that for both types of dynamics the
time-temperature superposition principle (TTSP) holds (see
Refs.~\cite{kob} and~\cite{gleim98}), i.e. that for a given dynamics
the shape of the correlation function does, in the $\alpha$-relaxation
regime, not depend on temperature. Therefore we conclude that not only
the temperature dependence of the time scale of the $\alpha$-relaxation
is independent of the microscopic dynamics but also those details of
the \mbox{($\alpha$-)}relaxation process that give rise to the stretching of
the correlation function. The same result is found for the coherent
intermediate scattering function and other values of $q$.

\begin{figure}[h]
\psfig{file=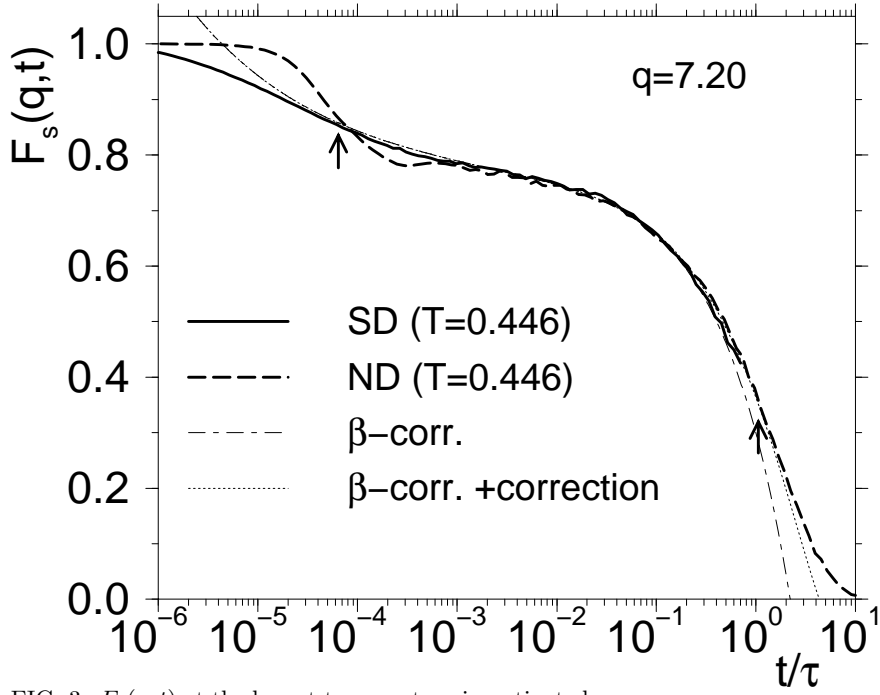,width=12cm,height=9.0cm}
\caption{$F_s(q,t)$ at the lowest temperature investigated versus
rescaled time $t/\tau(T)$ for the SD (bold solid line) and the ND (bold
dashed line). Dashed line: $\beta$-correlator for $\lambda=0.71$.  Thin
solid line: $\beta$-correlator + leading corrections at long times.}
\label{fig3}
\end{figure}

We now turn our attention to the $\beta$-relaxation, i.e. the
relaxation regime in which the correlation function is in the vicinity
of the plateau. From Fig.~\ref{fig2} we recognize that, for low
temperatures, the way the curves approach the plateau depends strongly
on the microscopic dynamics in that for the SD this approach is very
gentle whereas for the ND it is quite abrupt. This difference does not
exist for the {\it late} $\beta$-relaxation, i.e. when the correlation
functions start to fall under the plateau. This can be seen by plotting
the relaxation functions versus $t/\tau(T)$, which is done in
Fig.~\ref{fig3} for the SD and the ND for the lowest temperature
investigated. We see that this scaling of time makes the two
correlation functions fall on top of each other in the late
$\beta$-relaxation regime. 

MCT predicts that {\it in the $\beta$-relaxation regime} and
asymptotically close to the critical temperature $T_c$ the shape of the
correlation functions is given by the so-called $\beta$-correlator, a
function that can be computed within the framework of the
theory~\cite{mct1}. The form of this $\beta$-correlator depends on one
parameter, the so-called exponent parameter $\lambda$, which has been
computed for the present system to be $\lambda=0.71~\cite{nauroth97}$.
Using this theoretical value of $\lambda$ we fitted the correlation
curves in the $\beta$-regime with the corresponding $\beta$-correlator
and the resulting fit is show in Fig.~\ref{fig3}.  It has been shown
that, for temperatures slightly above $T_c$, the time window in which
the theoretical curve describes the data well can be extended
considerably, if also the {\it corrections} to the $\beta$-correlator
are taken into account~\cite{mct1,sciortino}.  For long times the
leading term of these corrections is given by a power-law with an
exponent which can be computed from $\lambda$.  The result of a fit to
our SD data with the $\beta$-correlator and the first correction at
long times is included in Fig.~\ref{fig3} as well. We recognize that
the fit with the $\beta$-correlator alone describes the SD data well
over about 3.5 decades in time. This time window is expanded at large
times by about half a decade by taking into account the corrections to
the $\beta$-correlator (time window between the two arrows). From the
figure we see that the theoretical curves fit the ND data only well in
the {\it late} $\beta$-relaxation regime, whereas they do a poor job in
the early $\beta$-relaxation regime since the asymptotic law are
completely obscured by the phonons~\cite{kob}. For the SD, however,
also the early $\beta$-relaxation is described very well by MCT in that
the approach to the (quasi) plateau is described well by the
$\beta$-correlator. The reason for this better agreement is likely that
in the SD the phonons are strongly damped and thus interfere much less
with the asymptotic laws of the theory on the time scale in which the
correlators approach this plateau.

\begin{figure}[h]
\psfig{file=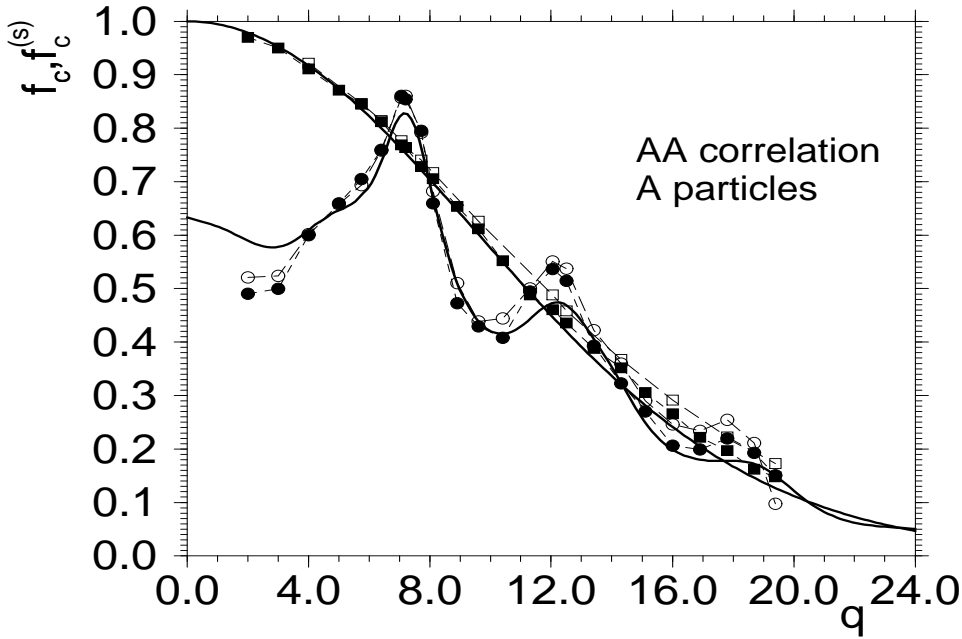,width=13cm,height=9.5cm}
\caption{Wave-vector dependence of the nonergodicity parameter for the SD
and ND (filled and closed symbols, respectively). The squares and circles
correspond to the incoherent and coherent intermediate scattering
function, respectively. The two solid lines are the prediction of MCT for
this system.}
\label{fig4}
\end{figure}

The last quantity we discuss is the $q$ dependence of $f_c$, the
so-called Edwards-Anderson or nonergodicity parameter. This quantity is
related to the height of the plateau in the correlation functions and
can be measured, e.g., in neutron scattering experiments (see, e.g.,
Ref.~\cite{frick90}). $f_c$ is one of the fit parameters for the
$\beta$-correlator and was thus obtained by performing such fits to the
incoherent and coherent intermediate scattering function at various
wave-vectors $q$. In Fig.~\ref{fig4} we show the $q$-dependence of the
corresponding nonergodicity parameters, $f_c^{(s)}(q)$ and $f_c(q)$,
for the SD and the ND for the coherent ($A-A$) and incoherent
intermediate ($A$) scattering function. We see that for all $q$
considered the two $f_c$ for the SD are very close to the ones for the
ND, thus showing that also this quantity does not depend on the
microscopic dynamics. Also included in the figure are the two curves
that correspond to the $q$-dependence of the $f_c$ as predicted by
MCT~\cite{nauroth97}. We see that these two theoretical curves agree
very well with the ones measured in the simulation. The reader should
appreciate that in the calculation of these theoretical curves {\it no}
free fit parameter was involved, since the only input to the MCT
calculations was the temperature dependence of the partial structure
factors, which were determined from the simulation. Thus we conclude
that the nonergodicity parameters do not depend on the microscopic
dynamics and can be calculated with high precision from MCT. (We also
mention that a comparison between the $q$-dependence of the
nonergodicity parameter of the ND and the ones predicted from MCT has
been made in Ref.~\cite{nauroth97}, using the $f_c(q)$ and $f_c^{(s)}(q)$
as determined in Ref.~\cite{kob}. However, as we found out in the course
of the present work, these $q$-dependencies were affected by a
systematic error which originated from the neglect of the corrections
of the $\beta$-correlator~\cite{mct1,sciortino}. Therefore the curves
for the ND shown in Fig.~\ref{fig4} are not quite the same as the ones
shown in Ref.~\cite{nauroth97}.)

In summary we conclude that at low temperatures the $\alpha$-relaxation
dynamics of this system is independent of the microscopic dynamics
whereas the early $\beta$-relaxation dynamics does depend on it. We
find that in the $\beta$-relaxation regime the relaxation behavior for
the SD is in very good agreement with the one predicted by MCT and that
this theory is also able to give a very accurate prediction of the
$q$-dependence of the nonergodicity parameters.

Acknowledgements: We thank W. G\"otze for valuable discussions and him
as well as H. L\"owen for constructive comments on a draft of this
paper. Part of this work was supported by the Deutsche
Forschungsgemeinschaft under SFB 262/D1.

\clearpage
\newpage
\clearpage
\newpage

\end{document}